\documentclass[aps,pra, twocolumn,groupaddress,twocolumn,10pt]{revtex4-1} 
\usepackage{amsmath}
\usepackage{latexsym}
\usepackage{graphicx}
\usepackage{amsfonts}
\usepackage{braket}
\usepackage{hyperref}
\usepackage{natbib}
\usepackage{amssymb}
\usepackage{nicefrac}
\usepackage{fancyhdr}
\usepackage[utf8]{inputenc}
\usepackage{dsfont}
\usepackage{colortbl}
\usepackage{xcolor}
\usepackage{subfigure}
\usepackage{psfrag}
\usepackage{tikz}
\usepackage{paralist}
\usepackage{nicefrac}
\usepackage{ulem}
\usetikzlibrary{arrows,shapes,decorations.pathmorphing}

\hypersetup{
    bookmarks=true,         
    bookmarksopen=true,			
    bookmarksopenlevel=1,		
    unicode=false,          
    pdftoolbar=true,        
    pdfmenubar=true,        
    pdffitwindow=false,     
    pdfstartview={FitH},    
    pdftitle={Simultaneous gates in frequency-crowded multi-level systems using fast, robust analytic control shapes},    
    pdfauthor={Lukas Theis},     
    colorlinks=true,       
    linkcolor=blue,          
    citecolor=blue,        
    urlcolor=blue           
}

	\newcommand{\comut}[2]{\left[ #1 , #2 \right]} 
	\newcommand{\op}[1]{\hat{\mathrm{#1}}}
	\newcommand{\id}{\mathds{\hat{1}}}
	
	\newcommand{\dt}{\mathrm{d}}
	\newcommand{\erf}[1]{\mathrm{erf}\left(#1\right)}
	
	\newcommand{\real}[1]{\operatorname{Re}\left\{#1\right\}} 
	\newcommand{\imag}[1]{\operatorname{Im}\left\{#1\right\}} 

	\newcommand{\del}{\mathrm{\Delta}}
	\newcommand{\ketbra}[2]{\ket{#1}\!\bra{#2}}
	\newcommand{\Ketbra}[2]{\Ket{#1}\!\Bra{#2}}
	\newcommand{\e}{\varepsilon}
	\newcommand{\LR}[1]{\left(#1\right)}
	\newcommand{\abs}[1]{\left\lvert #1 \right\rvert}
	\newcommand{\norm}[1]{\left\lVert #1 \right\rVert}

	\newcommand{\Trace}[1]{\text{Tr}\LR{#1}}		
	\newcommand{\PTrace}[2]{\text{Tr}_{#1}\left\{#2\right\}}	

	\newcommand{\refsec}[1]{Sec.\,\ref{sec:#1}}
	\newcommand{\reffig}[1]{Fig.\,\ref{fig:#1}}
	\newcommand{\refeq}[1]{(\ref{eq:#1})}
	\newcommand{\reftab}[1]{Tab.\,\ref{tab:#1}}

\makeindex

\begin{document}

\title{Simultaneous gates in frequency-crowded multilevel systems using fast, robust, analytic control shapes}

\author{L. S. Theis}
\email{luk@lusi.uni-sb.de}
\author{F. Motzoi}
\author{F. K. Wilhelm}
\affiliation{Theoretical Physics, Saarland University, 66123 Saarbr{\"u}cken, Germany}


\begin{abstract}
We present a few-parameter ansatz for pulses to implement a broad set of simultaneous single-qubit rotations in frequency--crowded multi-level systems. Specifically, we consider a system of two qutrits whose working and leakage transitions suffer from spectral crowding (detuned by $\delta$). In order to achieve precise controllability, we make use of two driving fields (each having two quadratures) at two different tones to simultaneously apply arbitrary combinations of rotations about axes in the X--Y plane to both qubits. Expanding the waveforms in terms of Hanning windows, we show how analytic pulses containing smooth and composite-pulse features can easily achieve gate errors $<10^{-4}$ and considerably outperform known adiabatic techniques. Moreover, we find a generalization of the WahWah method by Schutjens \emph{et al.} [Phys. Rev. A \textbf{88}, 052330 (2013)] that allows precise separate single-qubit rotations for all gate times beyond a quantum speed limit. We find in all cases a quantum speed limit slightly below $2\pi/\delta$ for the gate time and show that our pulses are robust against variations in system parameters and filtering due to transfer functions, making them suitable for experimental implementations.
\end{abstract}

\maketitle

\section{Introduction}
Quantum information processing and offshoots thereof, such as quantum computing and control are rapidly evolving fields in physics. Often, these involve physical systems for which a computational subspace can be singled out of a larger Hilbert space that is feasible for logical operations. Candidates for systems with such a structure are for instance several types of superconducting qubits \cite{Clark_Nature_453_1031}, optical lattices \cite{Maneshi_PRA_77_022303}, quantum dots \cite{Petta_Science_309_2180,Schliemann_PRB_63_085311}, Rydberg atoms \cite{Urban_NatPhys_5_110,Saffman_RevModPhys_82_2313}, neutral atoms \cite{Brickman_NJP_11_055022}, diamond NV centers \cite{Dobrovitski_AnnRevCondMatt_4_23}, trapped ions \cite{Monroe_PRA_89_022317,Haeffner_PhysRep_469_155} and nuclear magnetic resonance \cite{Jones_NMR_59_91}. A qubit subspace in these systems is typically formed by two levels of a multi-level system, such as an anharmonic ladder. This directly connects to leakage out of the computational subspace -- especially on short timescales -- since the transition frequencies of successive levels may sometimes only differ by a few percent \cite{Steffen_PRB_68_224518}.

We focus our attention on superconducting qubits embedded in circuit QED (cQED) architecture, where they are coupled to microwave cavities \cite{RevModPhys_73_357, You_PRB_68_064509, Schoelkopf_Nature_451_664}, both because it is a leading candidate for quantum computation, and because of the critical role that leakage and frequency-crowding play in such systems. Like the other candidate architectures, decoherence will set a timescale below which gates must be performed. Decoherence may for example arise from the electromagnetic environment \cite{Koch_PRA_76_042319} as well as from intrinsic material properties \cite{Martinis_PRL_95_210503}. Optimized designs of superconducting qubits, such as the transmon \cite{Koch_PRA_76_042319} and its 3D version \cite{Paik_PRL_107_240501, Rigetti_PRB_86_100506} avoid decoherence and allow for coherence times approaching 100$\mu$s, thus an improvement by several orders of magnitude compared to previous designs having coherence times of a few nanoseconds \cite{Schoelkopf_Nature_451_664}.

Experimental progress in implementing fast gates has been significant, achieving fidelities exceeding $99\%$ for both single-qubit and two-qubit operations \cite{Chow_PRL_102_090502, Barends_Nature_508_500}, and fidelities $\sim\! 85\%$ for two-qubit algorithms like Grover's search algorithm \cite{DiCarlo_Nature_460_240}. Leakage was removed in these experiments using the \textit{derivative removal by adiabatic gate} (DRAG) \cite{Motzoi_PRL_103_110501, Gambetta_PRA_83_012308, Motzoi_PRA_88_062318} technique, which works by considering the adiabatic effect of driving the leakage transition on the qubit's transition.

Apart from decoherence and leakage, addressability of single qubits in a multi-qubit layout \cite{Gershenfeld_Science_275_350, Pioro_NatPhys_4_776} is another key challenge for large scale quantum computing. Often enough it is impossible to have a single control field for each qubit, for instance if more than one qubit are in the same cavity, hence being spatially too close to be addressed individually \cite{Motzoi_PRA_88_062318}. Then, the applied field collectively drives all qubits at once, necessitating addressing qubit control by an internal parameter such as carrier frequency. A phenomenon that arises in such cases is frequency-crowding \cite{Schutjens_PRA_88_052330, Vesterinen_arXiv_1405.0450, Economou_PRB_91_161405}, e.g. describing a situation where logical transitions are well separated but for instance the working transition of one qubit is too close to a leakage transition of another qubit to be driven individually. In Ref. \cite{Schutjens_PRA_88_052330} a situation with two 3D transmons in one cavity was studied with respect to driving a rotation on one qubit, leaving alone the other one. They showed that it is possible to achieve a high fidelity by supplementing a Gaussian pulse with sideband modulation and a DRAG component, yielding pulse shapes named \textit{WahWah (Weak AnHarmonicity With Average Hamiltonian)}. Their results reveal that the WahWah ansatz works well if the targeted gate time is well-chosen.

A key requirement for quantum computing on a large scale is for errors arising from leakage, addressability, and frequency-crowding to fall well below a fault-tolerant threshold where quantum error correction (QEC) schemes \cite{Shor_PRA_52_R2493, Reed_Nature_482_382} improve with system size -- as has been recently demonstrated by Kelly \emph{et al.} \cite{Kelly_Nature_519_66} for a linear array of up to nine qubits. Conservative estimates \cite{Knill_Nature_434_39,Sanders_arXiv_1501.04932} for this threshold state that the average error per gate should be $\lesssim 10^{-4}$, with lower errors also requiring far fewer physical qubits per logical qubit, and with leakage out of the computational subspace being an additional but not insurmountable impediment\cite{Gosh_PRA_88_062329}. Thus, it is all the more pressing that these forms of error be suppressed.

These problems become compounded when multiple gates need to be operated simultaneously.  This operation is crucial to reducing computation times, which in turn allows for algorithm error linearly smaller in the number of qubits, as well as lower error budgets in QEC, due to error also accumulating during memory operations.  In the presence of near-resonances in crowded systems on the order of $1\%$ of the natural frequencies, pulses derived from adiabatic techniques are too short with respect to the decoherence time, whereas sideband modulation in the spirit of Ref. \cite{Schutjens_PRA_88_052330} disrupts concurrent gates with which it does not commute, leading to disruption of the desired final-time spectrum and even worse leakage than not including it.

In this work, we rather consider decoupling the effect of the leakage transition using interference between the different portions of the gate, being reminiscent of composite pulses. This is accomplished by expanding the control pulses in terms of (Hanning) windows, which include a third harmonic that can be used to drive three sequential rotations with alternating direction. The weights of the components are optimized numerically, yielding high-fidelity simultaneous X and/or Y gates with very short quantum speed limit, on the order of $2\pi/\delta$, where $\delta$ is the detuning between the two most closely crowded frequencies. The waveform shapes found are similar to the CORPSE \cite{Tycko_JChemPhys_83_2775} pulse sequence, and share a robustness to deterministic and non-deterministic phase shifts during the pulse.

This work is structured as follows: In \refsec{model_sys} we introduce the system composed of two (3D) transmons being collectively driven through a cavity. We derive various frame transformations that allow for analytically quantifying the error. Within \refsec{gates} we outline the analytic ansatz that is further optimized numerically and analyzed with respect to certain properties, such as quantum speed limit and robustness, within Sections \ref{sec:results} and \ref{sec:robustness}. In \refsec{shaping} we outline the degrees of freedom and show samples of our pulses. An extended model for WahWah pulses is given in \refsec{impWW}, achieving high fidelities for all gate times beyond a quantum speed limit.

\section{System}\label{sec:model_sys}%
\subsection{Model in the lab frame}\label{sec:model_lab}%
\begin{figure}[tb]
  \centering
  \includegraphics[width=.9\linewidth]{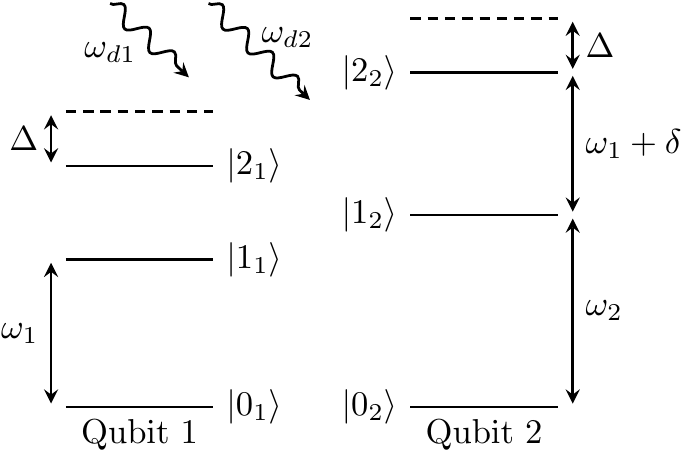}
  \caption{\label{fig:diagram}Energy level diagram of the two qubits. The leakage transition $\ket{1_2}\leftrightarrow\ket{2_2}$ of qubit 2 is only slightly detuned by $\delta$, which we call the crowding frequency in the remainder of this work, from the working transition $\ket{0_1}\leftrightarrow\ket{1_1}$ of qubit 1, i.e. $\omega_1+\delta = \omega_2 + \del$. The dashed lines indicate the level structure if the transmons where harmonic oscillators with frequencies $\omega_1$, $\omega_2$. We collectively drive the system with pulses at $\omega_{d1}$ and $\omega_{d2}$.}
\end{figure}

The computational subspace used for the qubit system in superconducting qubits is usually formed by the lowest two energy levels of an anharmonic oscillator with weak anharmonicity \cite{Koch_PRA_76_042319}, although this choice is not mandatory. In order to have a realistic model for the processes occurring in the system it is inevitable to consider at least the next higher energy level - which is referred to as a \textit{leakage level} in the remainder of this work. Starting in the lab frame we can break down the Hamiltonian of our system, which is composed of two superconducting transmon qubits, into a (constant) drift part $\op{H}_0$ and a controllable part $\op{H}_c$. Our bare two qubits are weakly anharmonic oscillators, described by $\op{H}_0$. Their energy level diagram is depicted  schematically in \reffig{diagram}. The lab Hamiltonian reads
\begin{align}\begin{split}\label{eq:system}
  \op{H} & = \op{H}_0 + \op{H}_c\\
  \op{H}_0 & = \sum\limits_{k=1}^{2}\left[\omega_k \op{n}_k + \del_k\op{\Pi}_2^{(k)}\right]\\
  \op{H}_c & = \Omega(t)\sum\limits_{j=1}^{2}\left[\lambda_j^{(1)}\op{\sigma}_{j,j-1}^{x(1)} + \lambda_j^{(2)}\op{\sigma}_{j,j-1}^{x(2)}\right].\end{split}
\end{align}
This form of the Hamiltonian can be derived from quantization of electrical circuits as shown in detail in Ref.~\cite{Bishop_PhD} and in Ref.~\cite{Blais_PRA_75_032329} for the two-level case. Given that the system is operated in the dispersive regime, i.e. cavity and qubit are sufficiently far detuned, the effective Hamiltonian can be recast into the form of Eq.~\refeq{system}. Without loss of generality, we can assume our anharmonicities to be equal, i.e. $\del_1=\del_2=\del$. This will not lead to any degeneracies since the detunings of different levels remain non-degenerate and independently tunable. The perpetually used projection and generalized Pauli operators are defined as
\begin{align}
  \op{\Pi}_j^{(k)} & = \ketbra{j}{j}^{(k)}\\
  \op{n}_k & = \sum\limits_j j\op{\Pi}_j^{(k)}\\
  \op{\sigma}_{j,j-1}^{x(k)} & = \Ketbra{j}{j-1}^{(k)}+\Ketbra{j-1}{j}^{(k)}\\
  \op{\sigma}_{j,j-1}^{y(k)} & = i\Ketbra{j}{j-1}^{(k)}-i\Ketbra{j-1}{j}^{(k)}
\end{align}
where the superscript $(k)$ refers to either qubit 1 or 2. $\Omega(t)$ is a semiclassical dipolar interaction control with four quadratures total, paired into two sets, whereby each set's carrier has frequency $\omega_{d1}$ and $\omega_{d2}$, respectively, so that
\begin{align}\label{eq:control_field}\begin{split}
  \Omega(t) & = \e_{x1}(t)\cos\left(\omega_{d1}t+\phi_1\right)+\e_{y1}(t)\sin\left(\omega_{d1}t+\phi_1\right)\\
& +\e_{x2}(t)\cos\left(\omega_{d2}t+\phi_2\right)+\e_{y2}(t)\sin\left(\omega_{d2}t+\phi_2\right).\end{split}
\end{align}
The $\lambda_j^{(k)}$ are associated to the relative strength at which the control field $\Omega(t)$ drives the leakage transition $\ket{1_k}\leftrightarrow\ket{2_k}$ compared to the working transitions $\ket{0_k}\leftrightarrow\ket{1_k}$. Numerical values for occurring frequencies and $\lambda_j^{(k)}$ are given in \reftab{systemparams} whereby the $\lambda_j^{(k)}$ are approximated to be $\sqrt{j}$ since the anharmonicities are assumed to be small (hence eigenstates are close to those of an harmonic oscillator) and since the transmons are far detuned from the driving resonator \cite{Gambetta_PRA_83_012308}. In principal, the relative phases $\phi_{1,2}$ between envelope and carrier need to be taken into account. However, since we apply the rotating wave approximation (RWA) throughout following sections, those phases become irrelevant \cite{Gambetta_PRA_83_012308, Motzoi_PRA_84_022307}. 
\begin{table}[htb]
  \setlength{\tabcolsep}{10pt}
  \centering
  \caption{\label{tab:systemparams}Definition of the system parameters\footnote[1]{Taken from Ref.\cite{Schutjens_PRA_88_052330}. Suggested for experiments by the group of Leo DiCarlo at TU Delft.}.}
  \begin{tabular}{lccl}
    \hline\hline
    & Qubit 1 & Qubit 2 & Unit\\
    $\omega_k/2\pi$ & 5.508 & 5.903 & GHz\\
    $\del /2\pi$ & -350 & -350 & MHz\\
    $\delta /2\pi$ & \multicolumn{2}{c}{45} & MHz\\
    $\lambda_1^{(k)}$ & 1 & 1 & \\
    $\lambda_2^{(k)}$ & $\sqrt{2}$ & $\sqrt{2}$ & \\
    \hline\hline
  \end{tabular}
\end{table}
\subsection{Frame transformations}\label{sec:trafos}%
Since we are interested in controllability of our system by microwave fields (amplitude and carrier) we choose to work in different frames, such as a frame rotating at a specific frequency in order to remove any intrinsic oscillation from the controls. Such a transformation to a frame $\ket{\Phi_R}=\op{R}^{\dagger}\ket{\Phi}$ rotating at frequencies $\omega_j^{(1,2)}$ is achieved by applying 
\begin{align}
  \op{R}(t) & = \left(\sum\limits_j e^{-i\omega_j^{(1)}t}\op{\Pi}_j^{(1)}\right) \otimes \left(\sum\limits_j e^{-i\omega_j^{(2)}t}\op{\Pi}_j^{(1)}\right)
\end{align}
according to the transformation rule
\begin{align}
  \op{H}^{R} & = \op{R}^{\dagger}\op{H}\op{R}+i\dot{\op{R}}^{\dagger}\op{R}. \label{eq:traforule}
\end{align}
Particular choices of $\omega_j^{(k)}$ can lead to frame representations with different meaning and application purposes, such as
\begin{align*}
  \begin{cases}
    \omega_j^{(k)} = j\omega_d & \text{, rotating frame with $\omega_d$}\\
    \omega_j^{(k)} = j\omega^{(k)}+\del_j^{(k)} & \text{, interaction picture}.
  \end{cases}
\end{align*}
Transforming to a different frame along the lines of Eq.\refeq{traforule} will also affect the unitary time evolution operator according to 
\begin{align}
  \op{U}^R(t) & = \op{R}(t)\op{U}(t)\op{R}^{\dagger}(0), \label{eq:trafotime}
\end{align}
where $\op{U}^R(t)$ is the time evolution operator inside the rotating frame.%
\subsubsection{Rotating frame}\label{sec:trafos_rot}%
In order to have a better insight in the dependence of our system on the two driving frequencies we choose to look at the rotating frame with $\omega_j^{(k)}=j\omega_{d1}$. For more controllability we pick our carrier frequencies with arbitrary fixed detuning $\Lambda_{1,2}$ from the qubits' resonance frequencies:
\begin{align}
  \omega_{d1} & = \omega_1 + \Lambda_1 \\
  \omega_{d2} & = \omega_2 + \Lambda_2
\end{align}
As $\op{H}_0$ and $\op{R}$ are diagonal, our new Hamiltonian reads 
\begin{align}
  \op{H}^R & = \op{H}_0 + \op{R}^{\dagger}\op{H}_c\op{R} + i\dot{\op{R}}^{\dagger}\op{R}.
\end{align}
In the remainder of this work we will use the shorthand notation
\begin{align}
  \Omega_j(t) & = \e_{xj}(t)+i\e_{yj}(t) \text{\quad and}\\
  \chi(t) & = \Omega_1(t)+e^{i\gamma t}\Omega_2(t),
\end{align}
with $\gamma = \omega_{d1}-\omega_{d2} = \del - \delta + \Lambda_1 - \Lambda_2$ being the detuning between both carrier tones. Performing the RWA, i.e. neglecting terms that oscillate with $\pm 2\omega_{d1,2}$ and $\pm \left(\omega_{d1}+\omega_{d2}\right)$, leads to the rotating frame Hamiltonian
\begin{align}\label{eq:ham_rot}
  \begin{split}
    \op{H}^R & = -\Lambda_1\op{\Pi}_1^{(1)} + (\del - 2\Lambda_1)\op{\Pi}_2^{(1)} \\
    & + (\delta-\del-\Lambda_1)\op{\Pi}_1^{(2)}+(2\delta-\del-2\Lambda_1)\op{\Pi}_2^{(2)}\\
    & + \left\{\frac{\chi(t)}{2}\sum\limits_{j=1}^2 \left[\lambda_j^{(1)}\ketbra{j}{j-1}^{(1)}+\lambda_j^{(2)}\ketbra{j}{j-1}^{(2)}\right]\right.\\
    & \left. + \vphantom{\sum\limits_{j=1}^2}\text{h.c.}\right\}.
  \end{split}
\end{align}%
Evidence for the validity of the RWA is that the system frequencies in the rotating frame are of the order of $\del$, thus more than an order of magnitude smaller than the qubit frequencies $\omega_{1,2}$. 
At this point we want to briefly comment on why a simple DRAG solution alone will in general not improve results. The main idea behind DRAG is to move to an adiabatic frame via a transformation $\op{V}$, satisfying $\op{V}(0)=\op{V}(t_g)=\id$. For simplicity we will restrict to only one driving field, hence equivalent to $\Omega_2(t)=\omega_{d2}=\Lambda_2=0$ in Eq.\refeq{ham_rot}. Essentially, the transformation matrix will then be
\begin{align}
	\op{V}(t) & =\exp\left(-i\frac{\e_{x1}(t)}{\kappa}\sum\limits_{j=1}^2 \left[\lambda_j^{(1)}\op{\sigma}_{j,j-1}^{y(1)}+\lambda_j^{(2)}\op{\sigma}_{j,j-1}^{y(2)}\right]\right).
\end{align}
A key to sufficiently simple DRAG solutions is fast convergence of a series expansion with respect to an expansion parameter $\eta = \e_{x1}/\kappa\ll1$ \cite{Motzoi_PRL_103_110501}. Performing the corresponding adiabatic expansion to first order will lead to multiple possible values for $\kappa$ and a compensation quadrature $\e_{y1} \propto (1/\kappa)\dot{\e}_{x1}$. For instance, $\kappa=\delta$ removes the leakage from qubit 2. However, this choice will leave in considerable phase error on both qubits and slightly increase errors from the other (less) crowded frequencies. In principle, similar problems occur for other choices of $\kappa$ so that a simple DRAG ansatz will not be able to effectively address frequency-crowding issues. One can also optimize within a set of higher derivative functions, as was done in Ref. \cite{Motzoi_PRA_88_062318}. However, we will show that a simpler non-adiabatic basis will exist for our problem.
\subsubsection{Interaction frame}\label{sec:trafos_int}%
Transforming the lab frame Hamiltonian according to Eq.\refeq{traforule} to the interaction frame yields the interaction Hamiltonian
\begin{align}\label{eq:ham_int}
  \begin{split}
    \op{H}_I & = \frac{\chi(t)}{2}\sum\limits_{j=1}^2\left[\lambda_j^{(1)}e^{i\delta_j^{(1)}t}\ketbra{j}{j-1}^{(1)}\right.\\
    & \left.+\lambda_j^{(2)}e^{i\delta_j^{(2)}t}\ketbra{j}{j-1}^{(2)}\right]\\
    & + \text{h.c.}
  \end{split}
\end{align}
where we have introduced the eigenfrequencies $\delta_1^{(1)} = -\Lambda_1$, $\delta_2^{(1)} = \del-\Lambda_1$, $\delta_1^{(2)} = \delta - \del -\Lambda_1$ and $\delta_2^{(2)} = \delta-\Lambda_1$. In the remainder of this paper we will be working in the frame defined by Eq.~\refeq{ham_int}.%
%
%
\section{Simultaneous single-qubit gates}\label{sec:gates}%
\subsection{Target evolutions}\label{sec:gates_target}
At first, we focus on implementing single qubit rotations about an rotation angle $\theta\in\mathds{C}$ so that in the lab frame the desired unitary for each qubit (reduced to the computational subspace) is given by
\begin{align}\label{eq:gate_rot}
	\op{U}_{\rm red}(t_g) & = e^{i\phi}\exp\left[-\frac{i}{2}\begin{pmatrix}0 & \theta\\ \theta^* & 0\end{pmatrix}\right],
\end{align}
with an arbitrary, yet unimportant global phase $\phi$. A complex rotation angle $\theta$ gives rise to arbitrary rotations around the X/Y axis. The real part $\real{\theta}$ will rotate around the X-axis whereas the imaginary part $\imag{\theta}$ rotates around the Y-axis. We will develop and design pulses starting from the interaction frame, so that throughout \refsec{magnus} the final time evolution operator of the two transmon system is approximately given by 
\begin{align}\label{eq:desgate}
  \op{U}(t_g) & = e^{i\phi}\bigotimes_{j=1}^{2}{\exp\left[-\frac{i}{2}\begin{pmatrix}0 & \theta_j & 0\\ \theta_j^* & 0 & 0\\0 & 0 & 0\end{pmatrix}\right]},
\end{align}
rotating qubit $j$ about an angle $\theta_j$ and neglecting phases on the leakage level. Consequently the gate we actually implement in the lab frame is obtained by utilizing Eq.\refeq{trafotime} to transform Eq.\refeq{desgate} back to the lab frame. Eventually this will lead to $\op{Z}$ errors (relative phase shifts) that are quantified in \refsec{phase_corrections} together with techniques to compensate them.
\subsection{Magnus expansion}\label{sec:magnus}%
\subsubsection{Basic idea}\label{sec:magnus_general}%
An arbitrary Hamiltonian $\op{H}(t)$ induces a time evolution after a time $t_g$ according to the time evolution operator
\begin{align}
  \op{U}(t_g) & = \mathcal{\hat{T}}\exp\left(-i\int\limits_0^{t_g}\dt t \op{H}(t)\right)
\end{align}
where $\mathcal{\hat{T}}$ is the time-ordering operator. The latter accounts for the general impossibility to compute a closed analytical form for $\op{U}(t)$ due to the fact that $\op{H}(t)$ in general does not commute with itself at different times. An approximation for the final unitary $\op{U}(t_g)$ can be expressed in terms of the Magnus expansion \cite{Magnus, Blanes_PhysRep_470_151}
\begin{align}
  \op{U}(t_g) & = \exp\left(-i\sum\limits_k\op{\Theta}_k(t_g)\right).
\end{align} 
The $\op{\Theta}_k$'s are hermitian matrices generated by nested time integrals over nested commutators of the Hamiltonian at different times. A major advantage of the Magnus expansion is the straightforward treatment  of time-ordering. The first two orders of the expansion are given by 
\begin{align}
  \begin{split}
    \op{\Theta}_0(t_g) & = \int\limits_0^{t_g}\dt t\op{H}(t),\\
    \op{\Theta}_1(t_g) & = -\frac{i}{2}\int\limits_0^{t_g}\dt t_2\int\limits_0^{t_2}\dt t_1\comut{\op{H}(t_2)}{\op{H}(t_1)}.
  \end{split}
\end{align}
Convergence of the series is not always guaranteed. For a differential equation $\dot{Y}(t)=A(t)Y(t)$ in a Hilbert space $\mathcal{H}$ with boundary condition $Y(0)=1$  it is proven \cite{Blanes_PhysRep_470_151} for bounded operators $A(t)$ that $\int_0^T\dt t\;\norm{A(t)} < \pi$ is a sufficient condition to guarantee convergence in the time interval $[0,T)$. Nevertheless, this condition is only a sufficient one so that the series may still converge for $t>T$ even if the previous criterion is not satisfied. In case of $Y(t)$ being a normal operator, in particular a unitary one, the series also converges for infinite dimensional problems.

\subsubsection{Lowest order conditions}\label{sec:magnus_conditions}%
Requiring the lowest order $\op{\Theta}_0(t_g)$ of the Magnus expansion to implement the gate of Eq.\refeq{gate_rot}, i.e. 
\begin{align}\label{eq:mag_final}
  \op{U}_F(t_g) = \exp\left(-i\op{\Theta}_0(t_g)\right)
\end{align}
gives a first idea about pulse characteristics. That way, errors may arise from higher orders in the expansion which can partially be related to an AC stark shift or a Bloch-Siegert shift \cite{Bloch_PR_57_522, Wei_JPhysB_30_4877}. Such errors can for instance be reduced by either requiring higher orders to vanish or to lead to global phase factors. Since the interaction Hamiltonian can be written as a direct sum $\op{H}_I=\op{H}_I^{(1)}\oplus\op{H}_I^{(2)}$ and using the identity $\exp(A\oplus B)=\exp(A)\otimes \exp(B)$ for $A=\op{H}_I^{(1)}$ and $B=\op{H}_I^{(2)}$ (superscript $(k)$ refers to qubit $k$) we find the conditions
\begin{subequations}\label{eq:conditions1}
  \begin{align}
    \lambda_1^{(1)}\int\limits_0^{t_g}\dt t\;e^{-i\Lambda_1 t}\chi(t) &= \theta_1 \label{eq:work1}\\
    \lambda_2^{(1)}\int\limits_0^{t_g}\dt t\; e^{i(\del-\Lambda_1) t}\chi(t) &= 0 \label{eq:leak1}\\
    \lambda_1^{(2)}\int\limits_0^{t_g}\dt t\; e^{i(\delta-\del-\Lambda_1) t}\chi(t) &= \theta_2 \label{eq:work2}\\
    \lambda_2^{(2)}\int\limits_0^{t_g}\dt t\; e^{i(\delta-\Lambda_1) t}\chi(t) &= 0 \label{eq:leak2}
  \end{align} 
\end{subequations}%
to implement unitaries having the form of Eq.\refeq{desgate}. Here, $\theta_{1,2} \in \mathds{C}$ are the rotation angles of qubit 1 and two, respectively. In the above equations the corresponding complex conjugate versions will also hold. For following calculations we introduce complex-valued amplitudes $a_{1,2}$ so that $\chi(t)=a_1\,\tilde{\Omega}_1(t)+a_2\,e^{i\gamma t}\tilde{\Omega}_2(t)$ with rescaled pulses $\tilde{\Omega}_{1,2}$. These rescaled pulses are further decomposed into their real and imaginary parts, i.e. $\tilde{\Omega}_j(t)=\tilde{\e}_{xj}(t)+i\tilde{\e}_{yj}(t)$. From Eq.\refeq{ham_int} it follows that Eqs.\refeq{work1} and \refeq{work2} are the working transitions of both qubits, whereas the other two equations belong to their leakage transitions. Semiclassically speaking, conditions \refeq{leak1} and \refeq{leak2} state that there must be no spectral power at the unwanted transitions whereas conditions \refeq{work1} and \refeq{work2} are instances of the area theorem. Terms arising from $\op{\Theta}_1(t_g)$ will mostly contribute to diagonal elements, thus characterizing $\op{Z}$ errors in the final unitary. Those $\op{Z}$ errors mainly stem from level shifts and population leaking out of the computational subspace while driving the system. %

\begin{figure*}[t]
  \centering
  \includegraphics[width=.9\linewidth]{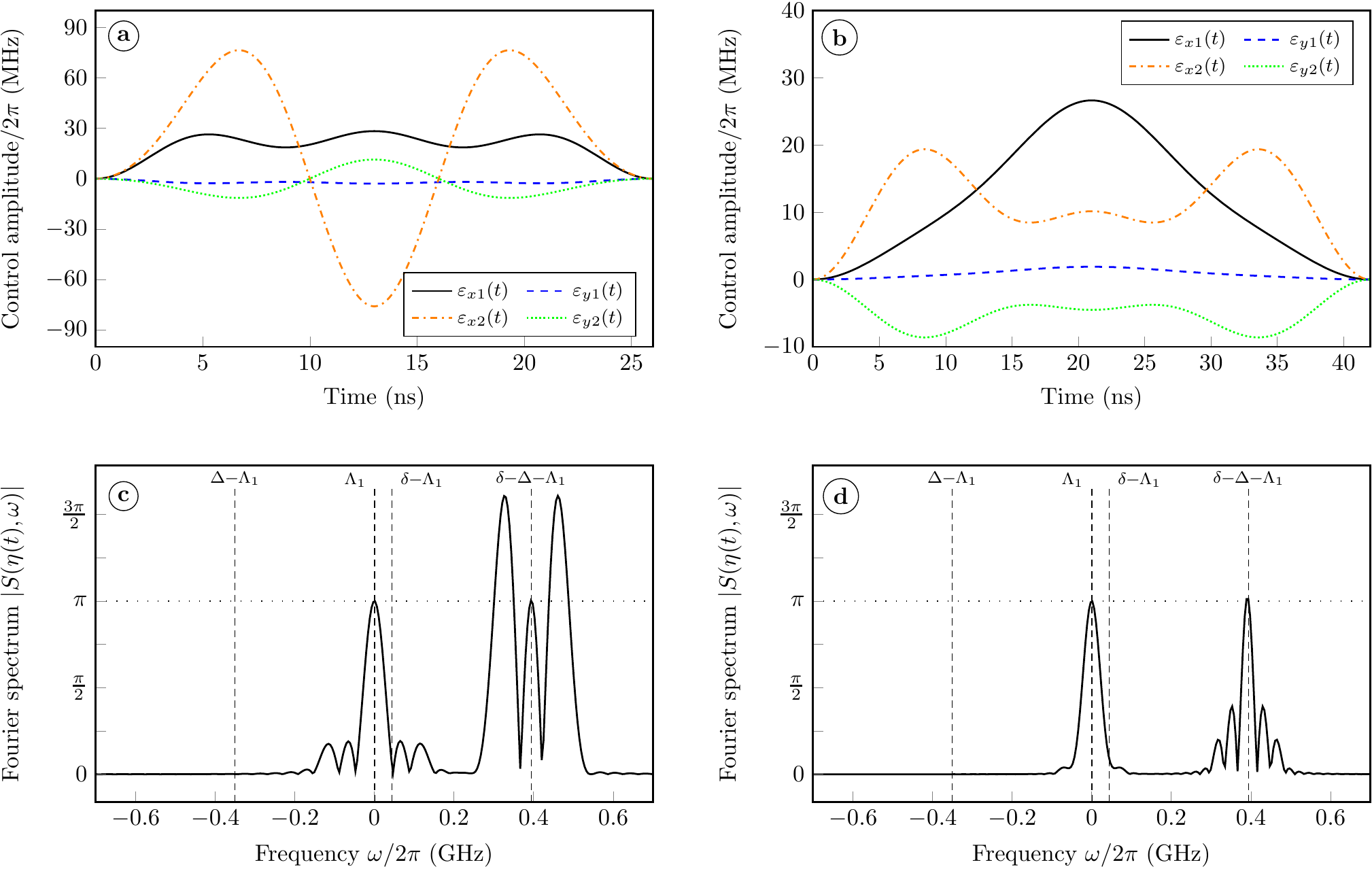}
  \caption{\label{fig:spectra_pulses}Control amplitudes for 26ns (a) and 42ns (b) together with their respective spectra (c,d). For longer gate times some controls resemble simple Gaussians, whereas they show characteristics of composite pulses especially at shorter gate times to account for large $\op{Z}$ errors during the sequence. The spectra indicate that Eqs.\refeq{conditions1} are essential to implement the desired rotations in crowded systems.}
\end{figure*}%

\subsection{Hanning windows}\label{sec:ansatz}
Gaussian pulse shapes are widely used to implement single-qubit rotations since they are smooth and have a limited bandwidth in excitation spectrum. Taking into account leakage levels and spectral crowding issues, new analytic methods such as DRAG and WahWah have been developed to counteract those challenges. However, in case of simultaneous gates, the large number of constraints makes the choice of basis functions (e.g. Hermite, Gaussian, Error function) especially important for considerations such as smoothness, boundary conditions and number of parameters.  

Yet, WahWah pulses only suppress leakage at the end of a gate, so that simply applying a WahWah control resonant with each transmon's logical transition will not allow simultaneous gates since there will be too much crosstalk between both physical qubits. We will choose to use the Hanning window functions \cite{Martinis_PRA_90_022307}
\begin{align}
  \Omega_{\rm H}(t) & = \sum\limits_{n=1}^N c_n\left[1-\cos\left(\frac{2\pi n t}{t_g}\right)\right], \label{eq:hanning}
\end{align}
which guarantees smoothness and also satisfies the boundary conditions for the pulse (start/end at zero). Utilizing this family of functions, we will also show that small errors at short times using only a small number of parameters can be obtained.

Hanning windows are expected to perform well in large part due to having both features that include smoothness (which helps to enforce adiabaticity) and composite-pulse structure that allows cancellation of errors between different (e.g. diabatic) parts of the pulse.  Moreover, the composite-pulse nature (very similar to the composite decoupling sequence, CORPSE \cite{Tycko_JChemPhys_83_2775}) 
can have intrinsic robustness to frequency miscalibrations, noise and time-dependent shifts (e.g. the Stark shift caused by off-resonant drive of the frequency crowded level), see \refsec{robustness}.

The performance of these pulses will be discussed and compared to Gaussians, derivative-based functions and the WahWah ansatz in the next sections.

\subsubsection{Constraints}\label{sec:shaping}%
From the lowest order Magnus expansion we find that the finite Fourier transform
\begin{align}
	S(\Omega,\rho) & = \int\limits_0^{t_g}\dt t\;\Omega(t) e^{i\rho t} \label{eq:fourier}
\end{align}
gives -- to lowest order -- a good intuition about the pulses' characteristics. Solving Eqs. \refeq{work1} and \refeq{work2} for the amplitudes $a_i$ so that the controls implement $\theta_i$-rotations yields
\begin{align}\label{eq:amplitudes_exact}\begin{split}
	a_1 & = \frac{\theta_1\left(\lambda_1^{(1)}\right)^{-1}-a_2 S(\tilde{\Omega}_2,\del-\delta-\Lambda_2)}{S(\tilde{\Omega}_1,-\Lambda_1)}\\
	a_2 & = \left(\frac{\theta_2}{\lambda_1^{(2)}}-\frac{\theta_1}{\lambda_1^{(1)}}\frac{S(\tilde{\Omega}_1,\delta-\del-\Lambda_1)}{S(\tilde{\Omega}_2,-\Lambda_1)}\right)\\\times & \left(S(\tilde{\Omega}_2,-\Lambda_2)-\frac{S(\tilde{\Omega}_1,\delta-\del-\Lambda_1)S(\tilde{\Omega}_2,\del-\delta-\Lambda_2)}{S(\tilde{\Omega}_1,-\Lambda_1)}\right)^{-1}.
\end{split}\end{align} 
Since the two logical transitions are detuned by $\delta-\del \sim 400\;{\rm MHz}$ from each other, it is valid to neglect the influence of both controls to the working transition they are not near-resonant with if their spectrum shows sufficient decay away from resonances, i.e. if for instance $S(\tilde{\Omega}_1,\delta-\del-\Lambda_1)/S(\tilde{\Omega}_2,-\Lambda_2)\ll 1$. Thus, we simply want each control to have an area complying with the desired angle of rotation $\theta_i$, hence
\begin{align}
  a_i & = \frac{\theta_i}{\lambda_1^{(i)}}\left(\int\limits_0^{t_g}\dt t\;\tilde{\Omega}_i(t)e^{-i\Lambda_it}\right)^{-1}. \label{eq:amplitude}
\end{align}
In fact, we observe that the approximation giving rise to Eq. \refeq{amplitude} leads to the same performance as the exact amplitude conditions in Eqs. \refeq{amplitudes_exact}. Finally, after picking an ansatz for the tilded x/y-control shapes, the controls in Eq.\refeq{control_field} are obtained as
\begin{equation}\label{eq:irradiate}
  \begin{alignedat}{2}
    \e_{x1}(t) & = \real{a_1\,\tilde{\Omega}_1(t)},\quad & \e_{y1}(t) & = \imag{a_1\,\tilde{\Omega}_1(t)} \\
    \e_{x2}(t) & = \real{a_2\,\tilde{\Omega}_2(t)},\quad & \e_{y2}(t) & = \imag{a_2\,\tilde{\Omega}_2(t)}.
  \end{alignedat}
\end{equation}

\section{Pulse Shaping\label{sec:shaping}}%

\subsection{Degrees of freedom \& Optimization\label{sec:optimization}}
To quantify our results we make use of the common overlap gate fidelity 
\begin{align}
  \Phi & = \frac{1}{d^2}\abs{\Trace{\op{U}_{\rm target}^{\dagger}\op{U}(t_g)}}^2 \label{eq:gate_fidelity}
\end{align}%
which is insensitive to global phases. Here, the trace is taken over the compound computational subspace $\left\{\ket{00},\ket{01},\ket{10},\ket{11}\right\}$ with $d=4$ being its dimension.

Investigating the performance of multiple ansatzes for the tilded controls reveals that it is satisfactory to choose $\tilde{\Omega}_{xj}$ to be a superposition of the first three Hanning windows, i.e. setting $N=3$ in Eq. \refeq{hanning}. Since one out of the three coefficients per control can be incorporated into the corresponding amplitude $a_i$, there are effectively four coefficients plus two detunings $\Lambda_{1,2}$ that determine the full set of parameters. To account for higher order errors, such as level shifts and $\op{Z}$ errors, we use the Nelder-Mead (NM) simplex algorithm \cite{CompJ_7_303} to obtain (locally) optimal values for the degrees of freedom. Since the tendency to get trapped in local extrema increases with the number of parameters \cite{Gao_ComputOptim_51_259} it is at least necessary to sample over various initial conditions and pick the best found solution. 

We can expect that generalization to more than two qubits if additional qubits do not add crowding on the order of $\delta$ is straightforward. In this case, all waveforms can be supplemented by DRAG to reduce the impact of crosstalk to other qubits -- scaling linearly with the number of qubits \cite{Motzoi_PRA_88_062318}. Even in case of multiple frequency-crowded qubits, one can in principle utilize higher order Hanning windows and/or additional quadratures to still achieve precise simultaneous gates. Nevertheless, this will ultimately increase the parameter space and thereby hamper the optimization. There are indications \cite{Tibbets_PRA_86_062309} that one might still observe reasonable convergence if operating far enough away from the quantum speed limit. A detailed study of this aspect is an area of future research.

\subsection{Sample pulses\label{sec:samplepulses}}

\begin{figure}[tb]
  \centering
  \includegraphics[width=.9\linewidth]{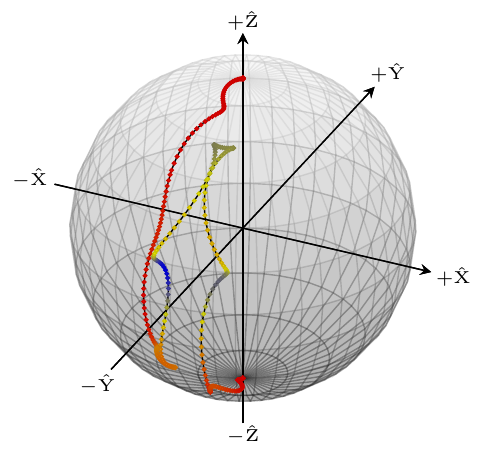}
  \caption{\label{fig:bloch}Visualization of the state evolution $\ket{\psi_2}$ with $\ket{\psi_2(0)}=\ket{0_2}$ for the 26ns pulses shown in \reffig{spectra_pulses}(a). Due to the composite-like structure $\op{Z}$ errors are corrected by approximately applying a $+2\pi/3\to -\pi/3\to +2\pi/3$ pulse. The color encodes the amount of population inside the computational subspace. Red means all population is inside $\{\ket{0},\ket{1}\}$ space, while blue indicates leakage by about $25\%$.}
\end{figure}

Examples of typical control shapes and their finite Fourier transforms are shown in \reffig{spectra_pulses}. For shorter gate times, i.e. 26ns in part (a), the $\e_{x2,y2}$ controls have magnitudes much larger than those required for a simple $\pi$ rotation. That is, these amplitudes push the pulses into the diabatic regime, with large intermediary leakage. Moreover, they show a composite-like structure very similar to a well-known method for robust pulses (CORPSE) \cite{Tycko_JChemPhys_83_2775}, approximately steering a $+2\pi/3\to -\pi/3\to +2\pi/3$ evolution on qubit 2, which is illustrated on the Bloch sphere in \reffig{bloch}. We will show in \refsec{robustness} that this composite structure enables robustness to frequency miscalibrations, noise and time-dependent shifts.

Our analysis shows that this composite-like shape, which is especially important for shorter gate times, is needed to achieve gate errors sufficiently small to make the pulses feasible for experimental implementations. Since leakage out of the computational subspace is not suppressed during the drive, there must be a precise interplay between $\e_{x1,y1}$ and $\e_{x2,y2}$ that ensures the correct ratio between how much population is pulled back into/out of the computational subspace by $\e_{x2,y2}$ while $\e_{x1,y1}$ drive a rotation on qubit 1 (still affecting the leakage transition on qubit 2 due to spectral crowding). Owing to severe leakage out of the $\{\ket{0_2},\ket{1_2}\}$ subspace and due to level shifts, there will be relative phase errors, manifesting as rotations around Z axis. This kind of error can be compensated by properly adjusting the interplay between all controls with the help of the NM algorithm. For longer gate times, such as 42ns in \reffig{spectra_pulses}(b), the composite structure is not that essential anymore: in this domain the type of control solutions becomes different. While $\e_{x1,y1}$ resemble simple Gaussian controls with maximal magnitude at $t=t_g/2$, their counterparts $\e_{x2,y2}$ are dominant at the beginning and end of the pulse. This also illustrates the interplay between all controls, leading to the correct cycling in and out of leakage subspace of qubit 2. The spectra in \reffig{spectra_pulses} illustrate that the lowest order conditions in Eqs.\refeq{conditions1} are essential to steer simultaneous rotations. Nevertheless, spectral arguments are not sufficient because of non-negligible higher order errors.

\section{Quantum speed limit\label{sec:results}}%

\subsection{Leakage error\label{sec:results_leakage}}
At first, we want to demonstrate that our method efficiently suppresses leakage arising from spectral crowding in our system. 
\begin{figure}[t]
  \centering
  \includegraphics[width=.9\linewidth]{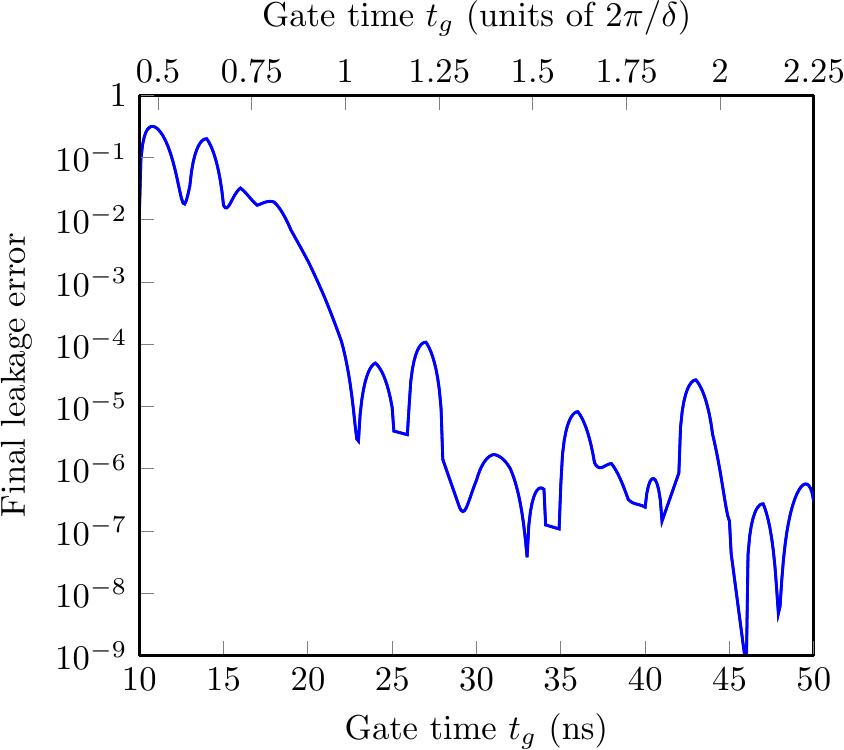}
  \caption{\label{fig:leakVSgate}Final leakage error of qubit 2 at different gate times for Hanning controls. Except for some gate times we achieve leakage errors below $10^{-5}$ after $\sim 25$ns. The non-monotone shape of the curve is most likely due to the system being overconstrained.}
\end{figure}
\reffig{leakVSgate} illustrates that we are able to achieve leakage errors (due to the critical $\ket{2_2}$ state) that are well below $10^{-5}$ after a gate time $\sim$ 25ns, indicating the relationship to the speed limit found in \refsec{limit}. The observed limit of approximately $25$ns for sufficiently suppressing leakage is largely explained by a bandwidth argument: For pulses of duration $t_g \lesssim 2\pi/\delta \sim 25$ns it is expected from spectroscopy that transitions with frequency close to $\delta$ will be excited, hence notable leakage out of the computational subspace occurs. Besides its impact on gate errors, leakage plays an important role in quantum error correction schemes. The majority of those assume that there will be no information leakage out of the computational subspace \cite{Gosh_PRA_88_062329}, making them vulnerable to leakage errors. 
\begin{figure}[b]
	\centering
	\includegraphics[width=.9\linewidth]{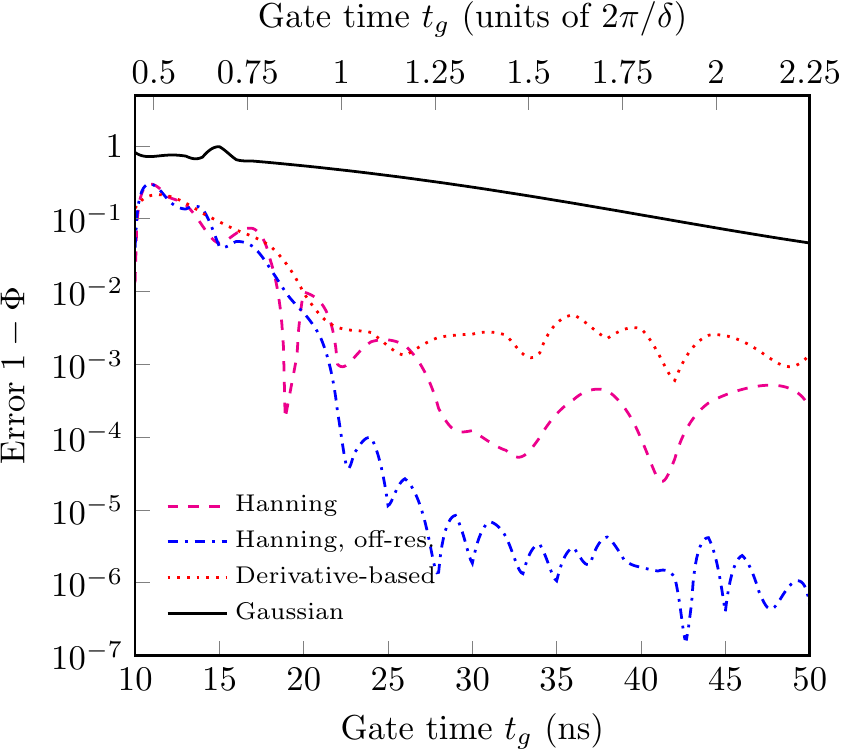}
	\caption{\label{fig:benchmark}Performance of Gaussian pulses (each resonant to $\omega_{1,2}$) , optimized Hanning-based pulses and derivative-based controls. Off-resonant controls yield an improvement by 2-3 orders of magnitude compared to their resonant counterparts.}
\end{figure}%

\subsection{Simultaneous rotations\label{sec:not_results}}
\begin{figure}[tb]
  \centering
  \includegraphics[width=.9\linewidth]{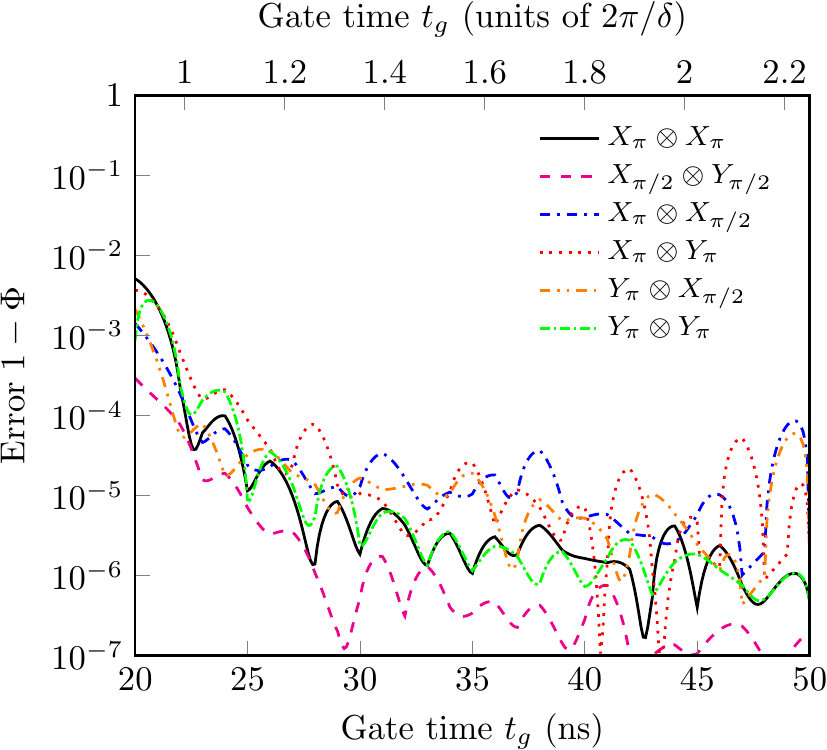}
  \caption{\label{fig:arbitrary_error}Average gate error as a function of gate time $t_g$ for various simultaneous rotations that are important for applications, for instance, in the AllXY sequence\cite{Reed_PhD_2013}.}
\end{figure}

 In \reffig{benchmark} we show performance of resonant (dashed magenta) and off-resonant (dash-dotted blue) Hanning controls together with a naive Gaussian approach, one per qubit (solid black), and a derivative-based ansatz (dotted red). Due to the finite bandwidth of the controls, at least the control resonant to $\omega_1$ will have severe spectral power at the leakage transition $\delta$ of the other qutrit, thus populating its $\ket{2}$ state and consequently leading to high gate errors in general. For instance, the gate error of Gaussian pulses slowly decreases exponentially with gate time. Derivative-based pulses perform roughly one order of magnitude worse compared to resonant Hanning controls, which is due to less control over phases via composite-like shapes. Clearly, using at least the first three Hanning windows ($N=3$) is sufficient for precise pulses in a frequency crowded system. Resonant controls enable average gate errors between $10^{-3}-10^{-4}$ while detuning the carrier slightly (by a few kHz to MHz) from the qubits' resonances further decreases the error by two orders of magnitude since there is considerably more control over level shifting effects (Stark, Bloch-Siegert) than in the resonant case. In fact, it may also be sufficient to apply only the controls $\e_{x2,y2}$ off-resonantly and leave $\e_{x1,y1}$ resonant with $\omega_1$: this will slightly increase the average gate error, however still being roughly 1.5 orders of magnitude better than having all controls on resonance. Nevertheless, the gain in fidelity by introducing more degrees of freedom comes at the cost of slower convergence, as has been stated in section \refsec{optimization}. Thus, fewer controls may be preferable in practical situations. \reffig{arbitrary_error} illustrates that our ansatz allows precise implementation of all simultaneous rotations around the X/Y axes. We focus on rotations that are part of the AllXY sequence \cite{Reed_PhD_2013} which can be used to tune up quantum systems and identify various error sources in order to underline the close connection between our results and recent experimental requests. Simultaneous gates involving identities are also easily implemented using Hanning-based pulses. However, in section \refsec{impWW} we present a generalization of Ref. \cite{Schutjens_PRA_88_052330} that is exactly designed for this purpose. Various other single qubit gates, such as the Hadamard gate 
\begin{align}\label{eq:hadamard}
  \op{U}_{\rm H} & = \frac{1}{\sqrt{2}}\begin{pmatrix}1 & \hphantom{-}1\\1 & -1\end{pmatrix}
\end{align}
can be decomposed into products of single qubit rotations. For instance, $\op{U}_{\rm H}=X_{\pi}Y_{\pi/2}$ (up to a global phase) may be implemented in roughly 40ns -- even in case of very strong crowding which is primarily analyzed in this work.

\subsection{Dependence on spectral detuning\label{sec:limit}}
\begin{figure}[tb]
  \centering
  \includegraphics[width=.9\linewidth]{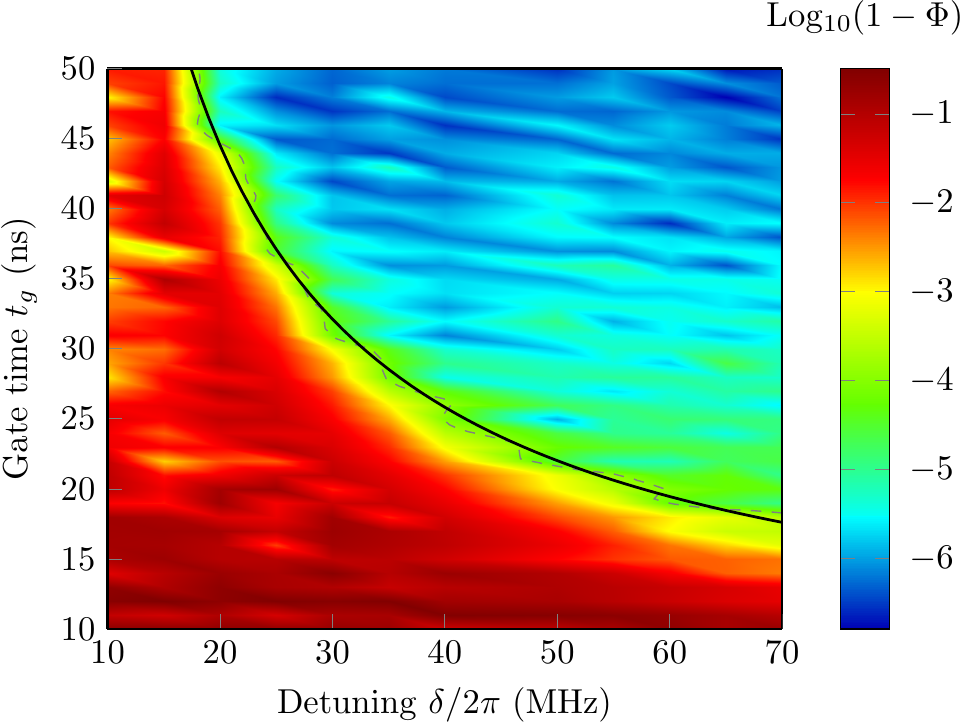}
  \caption{\label{fig:speedlimit}Average gate error as a function of gate time $t_g$ and crowding frequency $\delta$. The solid black line indicates the best fit for the speed limit (referring to $10^{-4}$ error) which is essentially $t_g^{\min}(\delta)\propto1/\delta$. This scaling indeed complies with the limit up to which leakage can be sufficiently be suppressed, as explained in \refsec{results_leakage}.}
\end{figure}

Because of the inevitable crosstalk between both transmon qubits, one expects a gate time (as a function of crowding $\delta$) at which precise control (we refer to the $10^{-4}$ error threshold) is no longer possible. Since excitations of transitions with frequency $\delta$ approximately scale $\propto 1/\delta$, we expect the quantum speed limit to scale in the same manner. In \reffig{speedlimit} we show the optimal average gate error as a function of gate time $t_g$ and crowding frequency $\delta$. The black line depicts the best fit, substantiating that the speed limit $t_g^{\min}(\delta)\sim 2\pi/\delta$.

\section{Separate single-qubit rotations \label{sec:impWW}}
As mentioned in the previous section, our Hanning-based pulses can also be utilized to implement $\id$ gates. Nevertheless, the WahWah technique \cite{Schutjens_PRA_88_052330} was designed for that purpose. There, only two control quadratures
\begin{subequations}\label{eq:wahwah_pulse}\begin{align}
	\e_x(t) & = A_{\pi}e^{-\frac{(t-t_g/2)^2}{2\sigma^2}}\left\{1-\cos\left(\omega_x \left(t-\frac{t_g}{2}\right)\right)\right\} \\
	\e_y(t) & = -\frac{1}{2\del}\dot{\e}_x(t)
\end{align}\end{subequations}
resonant with $\omega_1$ are exploited to drive a $\pi$-rotation on qubit 1 while leaving alone the second one, i.e. $\op{U}(t_g)=\op{X}\otimes\id$. To this end, a Gaussian is sideband-modulated with frequency $\omega_x$ that allows cancellation of leakage errors from qubit 2. The standard deviation is chosen as $\sigma=t_g/6$ to let the pulses smoothly start and end close to zero. In order to prevent qubit 1 from leaking, the in-phase control $\e_x$ is supplemented with DRAG whereby the factor of $2$ in the denominator stems from the absence of phase control \cite{Gambetta_PRA_83_012308}. 

\begin{figure}[tb]
  \centering
  \includegraphics[width=.9\linewidth]{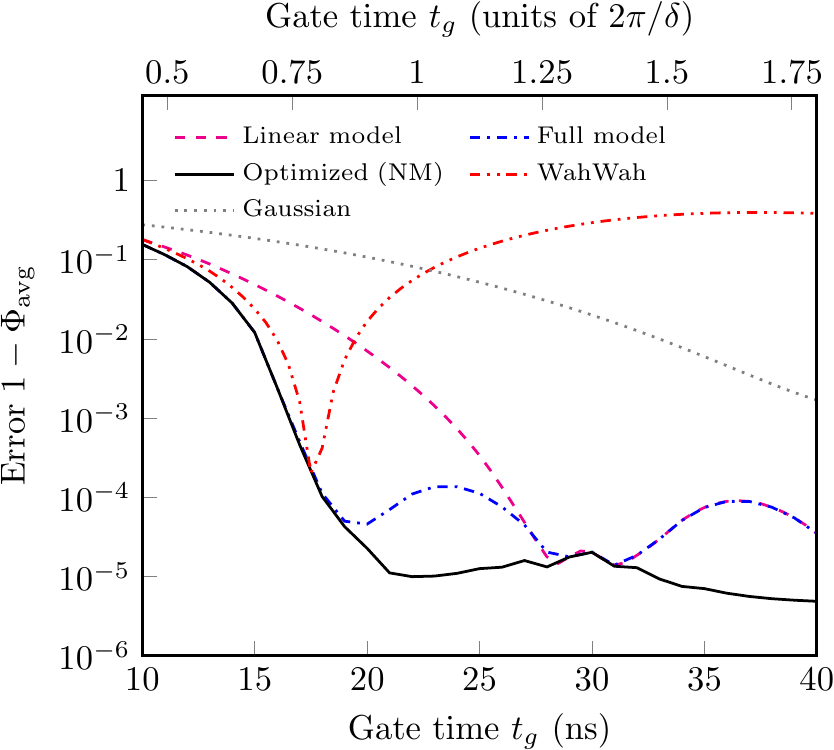}
  \caption{\label{fig:opt_wahwah}Gate error for numerically optimized WahWah pulses and pulses defined by the model in Eq.\refeq{wahwah_freq}, compared to a Gaussian and the original WahWah pulses. We observe a speed limit for the optimized pulses at $\sim 0.75~2\pi/\delta$.}
\end{figure}

As pointed out in Ref. \cite{Schutjens_PRA_88_052330}, there will be a relative phase error on qubit 2, eventually increasing the gate error measured via the fidelity in Eq.\refeq{gate_fidelity}. Therefore, the fidelity functions
\begin{align}\label{eq:gate_fidelity_nophase}
	\Phi_{\ket{*,i}} & = \frac{1}{4}\abs{\PTrace{\{\ket{0,i},\ket{1,i}\}}{\op{U}_F^{\dagger}\op{U}(t_g)}}^2\\
	\Phi_{\rm avg} & = \frac{1}{2}\left(\Phi_{\ket{*,0}}+\Phi_{\ket{*,1}}\right)
\end{align}  
are introduced. Here, the partial trace is taken over all states where the second qubit is either in $\ket{0}$ or $\ket{1}$. Hence, a high outcome of $\Phi_{\ket{*,i}}$ implies that qubit 2 starts and ends in state $\ket{i}$. Owing to this, the averaged reduced fidelity $\Phi_{\rm avg}$ is a way to estimate performance insensitive to relative phases in qubit 2. Originally, a modulation with $\omega_x=\delta/2$ was suggested and turned out to be successful if the gate time was well-chosen. From numerical studies, we derive the optimal sideband frequency $\bar{\omega}_x=\omega_x/\delta$ in terms of $\bar{t}_g=t_g\delta/2\pi$ as
\begin{align}\label{eq:wahwah_freq}\begin{split}
	\bar{\omega}_x(\bar{t}_g) & = \\
	 &\begin{cases}2.3~\erf{2.13\sqrt{\bar{t}_g-\frac{3}{4}}}, & \frac{3}{4} < \bar{t}_g \leq \frac{5}{4}\\
	2.3~\erf{\frac{2.13}{\sqrt{2}}}+0.41\left(\bar{t}_g-\frac{5}{4}\right),& \bar{t}_g > \frac{5}{4}.\end{cases}
\end{split}\end{align}
The model works well in a large range of $\delta/\del$ and incorporates a speed limit for the ansatz in Eqs.\refeq{wahwah_pulse}, which is approximately given by $0.75~2\pi/\delta$. For gate times close to the speed limit, there is a nonlinear dependence proportional to the error function $\erf{x}$. After passing through the nonlinear region $\bar{t}_g \gtrsim 5/4$ the sideband frequency scales linearly with the gate time. In \reffig{opt_wahwah} we illustrate performance of fully optimized pulses (solid black line), the piecewise model in Eq.\refeq{wahwah_freq} (dash-dotted blue line), the pure linear model for $\bar{\omega}_x$ (dashed magenta line) and a naive Gaussian approach (dotted gray line).

Phase errors in qubit 2 can be tracked and corrected by methods presented in \refsec{phase_corrections}.
\section{Robustness analysis\label{sec:robustness}}
An important property for experimental application of control pulses is their robustness against mismatches in for instance carrier frequencies or pulse envelopes, as well as filtering effects due to experimental hardware. One means to account for such imprecise knowledge is the Ad-HOC protocol \cite{Egger_PRL_112_240503} which combines open- and closed-loop optimal control, enabling enhancements in gate fidelities by an order of magnitude. However, since errors can in general occur during/in between operations of the hardware, e.g. $1/f$ noise, it is desired to work with pulses that show intrinsic robustness against certain errors and thereby go beyond closed-loop control methods. 
\subsection{Filtering effects}
\begin{figure}[tb]
   \centering
   \includegraphics[width=.9\linewidth]{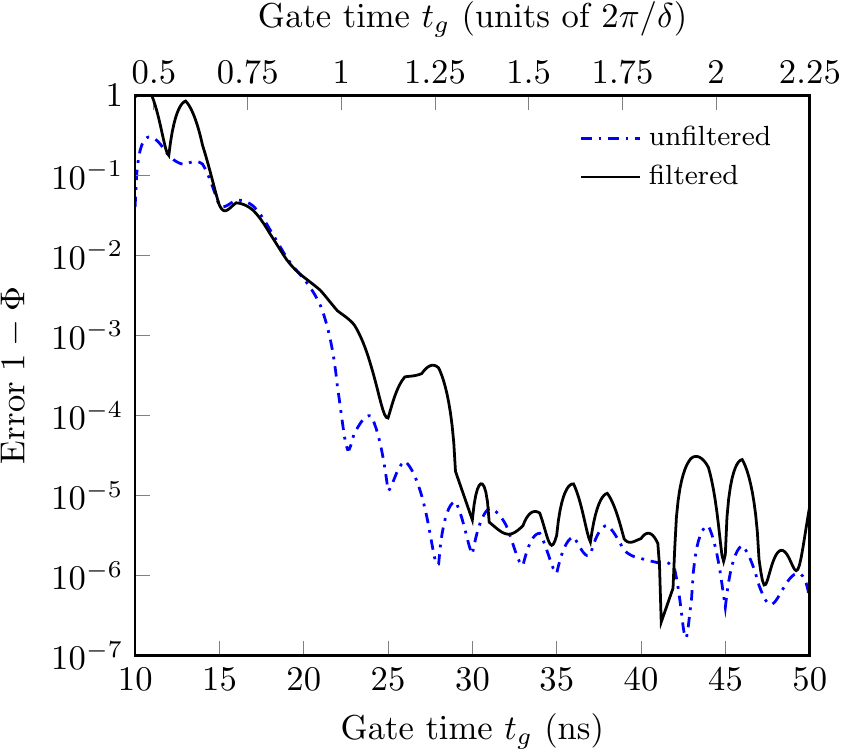}
   \caption{\label{fig:filtering}Performance of optimal control shapes under Gaussian filtering compared to the unfiltered case. Except from some gate times, almost all increase of error can be compensated by adjusting the pulse's magnitudes.}
 \end{figure}

When applying optimal control pulses in an experiment, their shape will be altered according to the transfer function of used hardware. For instance filtering effects due to finite bandwidth of waveform generators is a crucial part of those transfer functions. By transforming the controls in Eq.\refeq{irradiate} according to 
\begin{align}\label{eq:filter}
 u_{\rm filt.}(t) & = \frac{1}{2\pi}\int\limits_{-\infty}^{+\infty}\dt t'\int\limits_{-\infty}^{+\infty}\dt \omega\; F(\omega)e^{i(t-t')\omega}u_{\rm unfilt.}(t')
\end{align}
we model hardware filters. Here, $F(\omega)$ is the response function of the filter which is assumed to be Gaussian, i.e. $F(\omega)=\exp(-\omega^2/\omega_0^2)$ whereby $\omega_0/2\pi=425.4\;{\rm MHz}$ (approximation for  Tektronix AWG5014 \cite{Motzoi_PRA_84_022307}). Eq.\refeq{filter} describes a non-causal filter, which was found \cite{Motzoi_PRA_84_022307} to be a better approximation than a causal one. Applying this filter without further corrections has -- up to some exceptions -- no tremendous effect on fidelity, as shown in \reffig{filtering}. We find that fine-tuning the control amplitudes' magnitudes almost corrects for all additional error, either by precomputing the effect of the filter, or making use of closed-loop (Ad-HOC) type optimization in experiment. In general, if the transfer function of experimental hardware is known, its inverse can be incorporated into the input controls to achieve better performances\cite{Motzoi_PRA_84_022307}.
\subsection{System parameters}
\begin{figure}[tb]
  \centering
  \includegraphics[width=.9\linewidth]{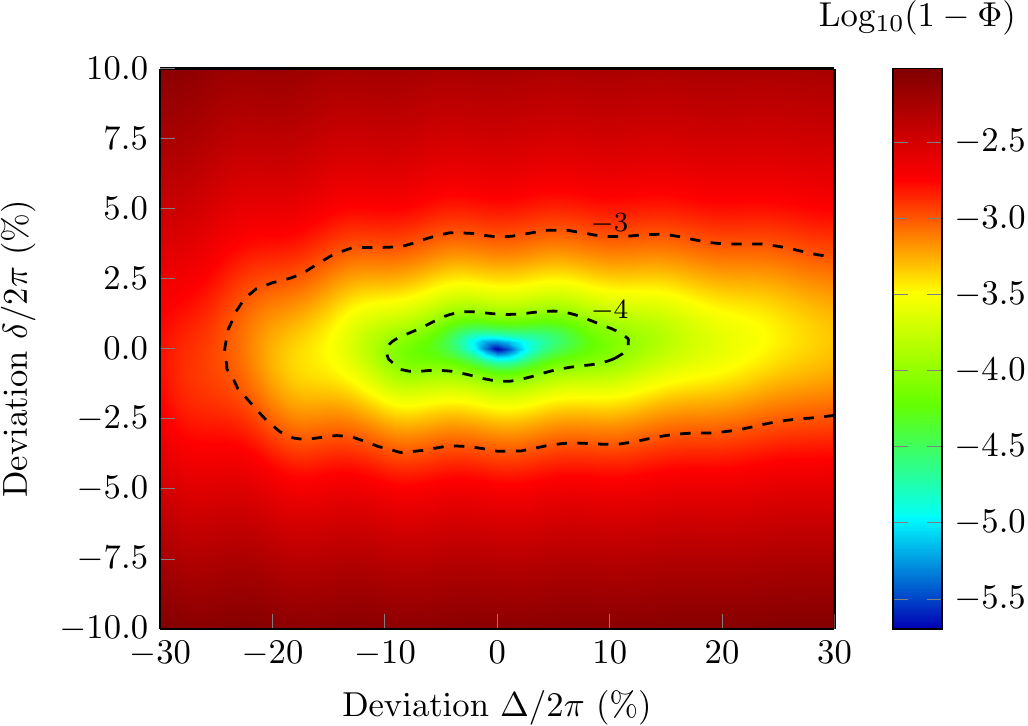}
  \caption{\label{fig:robustness_delDel}Error landscape for a $\op{X}\otimes\op{X}$ gate at 30ns. The landscape illustrates that our pulses are very robust against deviations of the anharmonicity $\del$. Although $\delta$ is expected to be the most crucial parameter, we are able to achieve errors $<10^{-3}$ if $\delta$ is known up to only $\pm 4\%$.}
\end{figure}

Besides filtering, also uncertainties in system parameters such as qubit frequencies may give rise to additional errors. The two most important frequencies being crucial for the system dynamics are the anharmonicity $\del$ and most notably the crowding frequency $\delta$. It is expected that the optimal solutions will strongly depend on those two frequencies, giving rise to the question about robustness of the solutions against imperfect knowledge of $\delta$ and $\del$. \reffig{robustness_delDel} depicts the average gate error under variations of $\del$ and $\delta$, showing that the solutions are very robust against mismatch in anharmonicity and pretty robust against uncertainties in $\delta$: Deviations of $\delta$ by $\pm 1.5\%$ and $\pm 4\%$ can still lead to gate errors of $10^{-4}$ and $10^{-3}$, respectively. The observed asymmetry in $\del$ stems from the fact that $\del < 0$. A negative deviation of $\del$ means that $\abs{\del}$ decreases, so that leakage errors inside qubit 1 become more likely whereas a positive deviation increases the distance between $\ket{1}\leftrightarrow\ket{2}$. 
Note that in general, also timing jitter from the arbitrary waveform generator will affect system performence. As it will be small and can by rescaling usually be compared to system parameter uncertainty, we expect its impact to be similarly small. 

\section{Correcting relative phases}\label{sec:phase_corrections}
As mentioned before (see \refsec{gates_target}) we introduce $\op{Z}$ errors and global phases in our lab frame gates by the ansatz we choose. Since global phases are irrelevant we do not have to account for them; however, $\op{Z}$ errors should be corrected. We will only focus on the correction inside the computational subspace because phases on the leakage levels are unimportant. Neglecting global phases, we calculate the $\op{Z}$ errors we make from Eq.\refeq{trafotime} as
\begin{align}
  \exp\left[i\frac{(\omega_{d1}-\Lambda_1)t_g}{2}\op{Z}\right]\otimes\exp\left[i\frac{(\omega_{d1}+\delta-\del-\Lambda_1)t_g}{2}\op{Z}\right].
\end{align}
This can easily be corrected if a $\op{Z}$ control is available on both qubits by applying controls $Z_1$ and $Z_2$ for a duration $T$ so that their areas are
\begin{align}
  \int_0^T \dt t\; Z_1(t) & = -\frac{(\omega_{d1}-\Lambda_1)t_g}{2},\\
  \int_0^T \dt t\; Z_2(t) & = -\frac{(\omega_{d1}+\delta-\del-\Lambda_1)t_g}{2}.
\end{align}

Alternatively it is also possible to account for this error by properly adjusting the phase of the preceding gate. This procedure is similar to what is named phase ramping \cite{Patt_JMR_96_94, Gambetta_PRA_83_012308} or frame compensation \cite{Motzoi_PRA_88_062318}. Essentially the XY-plane is rotated by an appropriate angle to account for the error made.
%
%
\section{Conclusions}
We have investigated independent control of two transmon qubits coupled to the same cavity. Typically, high fidelities in a system like this are hindered by spectral crowding, where for instance a harmful leakage transition is close in frequency to a logical transition (detuning $\delta$) and thereby renders individual addressability impossible. Our ansatz, based on a superposition of Hanning windows, achieves gate errors well below $10^{-4}$ while being subject to a quantum speed limit slightly below $2\pi/\delta$. We have shown that detuning the controls from resonance by a few kHz to MHz can again decrease the gate error by two orders of magnitude.

Our pulses outperform adiabatic methods and WahWah pulses for fast and precise implementations of simultaneous single-qubit rotations. Moreover, the ansatz theoretically allows for an easy generalization to more than 2 qubits, especially if additional ones do not add crowding on the order of $\delta$. We have adressed possible limitations in \refsec{optimization}.

Additionally, we have presented a model that generalizes WahWah pulses to arbitrary gate times and crowding frequencies, as long as not aiming at times below the quantum speed limit $\sim 0.75~2\pi/\delta$.

We have shown that pulse oscillations occur on timescales that are unsusceptible to filtering effects and that their composite-like shape gives rise to robustness against off-resonance errors, such as mismatches in characteristic frequencies. Moreover, the pulses are analytical and allow for easier debugging and benchmarking compared to fully numeric control shapes.
%
%
\section{Acknowledgements}
We would like to thank Leo DiCarlo and his group for suggesting this problem. This work was funded by the European Union within SCALEQIT.

\bibliography{Bibliography.bib}
\bibliographystyle{apsrev4-1}

\end{document}